\documentclass[twoside]{reporteq}
\usepackage{svcon2e}
\usepackage{makeidx}

\usepackage{latexsym}
\usepackage{amsmath}   
\usepackage{amsfonts}
\usepackage{amssymb}
\usepackage{graphicx}
\usepackage{amsthm}
\newcommand{\ed}{\end{document}}
\makeindex

\newcommand{\Lie}{\pounds}

\begin{document}

\chapter{Spinor Formulations for\break
Gravitational Energy-Momentum}
\chapterauthors{Chiang-Mei Chen, James M. Nester, and Roh-Suan Tung}
{ {\renewcommand{\thefootnote}{\fnsymbol{footnote}}
\footnotetext{\kern-15.3pt AMS Subject Classification: 83C05,
83C40, 83C60.} }}



\begin{abstract}
We first describe a class of spinor-curvature identities (SCI)
which have gravitational applications.  Then we sketch the topic
of gravitational energy-momentum, its connection with Hamiltonian
boundary terms and the issues of positivity and
(quasi)localization. Using certain SCIs several spinor expressions
for the Hamiltonian have been constructed.  One SCI leads to the
celebrated Witten positive energy proof and the Dougan-Mason
quasilocalization.  We found two other SCIs which give alternate
positive energy proofs and quasilocalizations. In each case the
spinor field has a different role.  These neat expressions for
gravitational energy-momentum have much appeal. However it seems
that such spinor formulations just have no room for angular
momentum; which leads us to doubt that spinor formulations can
really correctly capture the elusive gravitational
energy-momentum.\\
\noindent {\bf Keywords: } Gravitation, energy-momentum, positive
energy, quasi-local quantity, Hamiltonian.\par
\end{abstract}

\pagestyle{myheadings} \markboth{Chiang-Mei Chen, James M. Nester,
Roh-Suan Tung}{Spinor Formulations for Gravitational
Energy-Momentum}


\section{Introduction}

One of the most outstanding results in classical gravitation
theory (more specifically we mean in GR: general relativity,
Einstein's gravity theory), obtained via Clifford algebra and
spinor methods, was Witten's positive energy proof \cite{Wit81}.
This seminal work (which was inspired by an analogous result in
quantum supergravity \cite{DT77,Gri78}) led to many new ideas
regarding gravitational energy and its localization.  To
appreciate this work and its importance we need to recall some
facts about one of nature's most elusive quantities: gravitational
energy.

A suitable expression which could provide a physically reasonable
description of the energy-momentum density for gravitating systems has
long been sought.  All candidates had several shortcomings.  In
particular they violated a fundamental theoretical requirement---that
gravitational energy should be positive---as well as requirements
concerning localization and reference frame independence.  Since
Witten's positive energy proof (unlike the earlier indirect proof of
Schoen and Yau \cite{SY79}) can be understood in terms of the
Hamiltonian, the Hamiltonian density associated with this proof
provides a {\it locally positive localization}---and thus has real
promise as a {\it truly physical} energy-momentum density for
gravitational fields.  To fulfill this promise certain features need
further consideration; an outstanding one concerns the role of---and
even the need for---the spinor field, which seemed rather mysterious.

Here we give a survey (further details can be found in the
references) of three (classical, commuting, non-supersymmetric)
spinor/Clifford algebra formulations of the GR Hamiltonian known
to us.  We examine the underlying mathematics (spinor-curvature
identities), outline their associated positive energy proofs and
energy-momentum quasilocalizations, and note the various distinct
roles of the spinor field.  Although they give good expressions
for energy-momentum, these spinor expressions do not seem to have
to the proper qualities for giving a good description of
relativistic angular momentum.  Hence they apparently do not
succeed in giving a full physical description for the
energy-momentum density of asymptotically flat gravitating
systems.

The plan of this work is as follows.  In section 2 we present our
notation and conventions for geometric objects, forms, the Dirac
algebra, spinors, and ``Clifforms''.  In section 3 a succinct
presentation of a class of {\it spinor curvature identities} is
given; three special cases which have gravitational applications
are noted. Then in section 4 the topic of gravitational
energy-momentum is discussed; we note the fundamental theoretical
requirement of positivity (finally proved over 20 years ago) and
the still outstanding issue of the localization (or
quasi-localization) of gravitational energy-momentum.  In section
5 we explain the Hamiltonian approach to energy-momentum, noting
the important roles played by the Hamiltonian boundary term; the
standard ``ADM'' Hamiltonian (albeit in non-standard variables)
for GR along with a good choice of boundary term (which
necessarily requires reference values) are presented.  In section
6 we consider three alternate GR Hamiltonians obtained via certain
spinor curvature identities along with suitable associated spinor
field reparameterizations.  The first leads to the celebrated
Witten positive energy proof; it also gives the famous
Dougan-Mason energy-momentum quasilocalization.  The second
alternate Hamiltonian uses SU(2) spinors (i.e., the spinors of
3-dimensional space) in a way similar to, but distinct from, the
first case; it yields another proof and quasilocalization.  The
third alternative is fundamentally quite different; unlike the
former cases (where the spinor field was introduced as an
alternate reparameterization of the Hamiltonian) now the spinor
field enters into the Lagrangian as a dynamic physical field; this
case yields yet another positivity proof and quasilocalization.
The following section 7 notes that (i) the spinor formulations
have the extremely nice property of not requiring any explicit
reference values, (ii) in all cases the {\it Hamiltonian boundary
variation principle} reveals the associated boundary conditions,
and (iii) the various distinct roles played by the spinor fields
in these three formulations.  In section 8 consideration is given
to the other conserved quantities of an asymptotically flat space:
angular momentum and the center-of-mass moment.  The key role
played by a certain type of term in the conventional Hamiltonian
and the absence of this type of term in all of the aforementioned
spinor formulations is noted.  The concluding section summarizes
the virtues of the spinor Hamiltonian expressions; we note,
however, that the limitation in connection with angular momentum
and especially the center-of-mass moment raises grave doubts as to
whether these spinor formulations are really properly representing
the physics of gravitating systems.

\bigskip

\section{Conventions}

\medskip
Geometry (especially with a metric compatible connection, as is
assumed here) can be conveniently described using differential
forms.  The basis one-forms $\vartheta^\mu:=e^\mu{}_i dx^i$ (dual
to the basis vectors $e_\nu$) are chosen to be orthonormal (and
are generally non-holonomic).  The metric is then given by
$g=g_{\mu\nu}\vartheta^\mu\otimes\vartheta^\nu$  with $g_{\mu\nu}$
constant.  (We will not need the coordinate frame metric
components $g_{ij}=g_{\mu\nu}e^\mu{}_i e^\nu{}_j .$)  The
connection is described by the one-form
$\Gamma^{\mu\nu}=\Gamma^{\mu\nu}{}_i dx^i .$  Because of the
metric compatibility condition (and the the frame type choice) the
connection one-form is antisymmetric:
$\Gamma^{\mu\nu}\equiv\Gamma^{[\mu\nu]} .$ In general the
connection may have torsion, which is neatly described by the
2-form field
\begin{equation}
T^\mu:=D\vartheta^\mu:=
d\vartheta^\mu+\Gamma^\mu{}_\nu\wedge\vartheta^\nu
=(1/2)T^\mu{}_{\alpha\beta}\vartheta^\alpha\wedge\vartheta^\beta.
\end{equation}
The curvature is also described by a 2-form field:
\begin{equation}
R^\alpha{}_\beta:=d\Gamma^\alpha{}_\beta+\Gamma^\alpha{}_\gamma\wedge
 \Gamma^\gamma{}_\beta
=(1/2)R^\alpha{}_{\beta\mu\nu}\vartheta^\mu\wedge\vartheta^\nu.
\end{equation}
The unit volume element (in 4-dimensional spacetime) is given by
$\eta:=\vartheta^0\wedge\vartheta^1\wedge\vartheta^2\wedge\vartheta^3$.
We often use the dual Grassmann basis, which can be constructed
from $\eta$ by contraction (aka the interior product):
$\eta_\mu:=i_{e_\mu}\eta$, $\eta_{\mu\nu}:=i_{e_\mu}\eta_\nu$,
$\eta_{\mu\nu\alpha}:=i_{e_\mu}\eta_{\nu\alpha}$, and
$\eta_{\mu\nu\alpha\beta}:=i_{e_\mu}\eta_{\nu\alpha\beta}$, the
latter is the totally antisymmetric Levi-Civita tensor.
Alternately these objects can be obtained via the Hodge dual:
$\eta^{\mu\nu\dots}:=*(\vartheta^\mu\wedge\vartheta^\nu\wedge\
\dots$). A succinct notation, neatly suited to our material, is
Geometric (Clifford) Algebra valued forms, sometimes referred to
as {\it Clifforms} \index{Clifforms} \cite{DMH91,Mi87,Est91}.
With the Dirac conventions
\begin{equation}
\gamma_\alpha\gamma_\beta+\gamma_\beta\gamma_\alpha=
2g_{\alpha\beta} = \hbox{diag}(+1,-1,-1,-1),
\end{equation}
\begin{equation}
\gamma_{\alpha\beta\dots}:=\gamma_{[\alpha}\gamma_\beta\dots{}_{]},
\quad \gamma:=\gamma^0\gamma^1\gamma^2\gamma^3,
\end{equation}
\begin{equation}
\bar\psi:=\psi^{\dag}\beta, \qquad \beta\equiv\beta^{\dag}, \qquad
(\beta\gamma^\mu)^{\dag}\equiv \beta\gamma^\mu,
\end{equation}
we define the {\it frame}
(a vector valued one-form),
 the {\it torsion} \index{torsion} (a vector valued 2-form),
 the {\it connection} \index{curvature} (a bivector valued
one-form),
 and the
{\it curvature} (a bivector valued 2-form), respectively, by
\begin{align}
\vartheta:=\vartheta^\mu\gamma_\mu, &\quad
T:=D\vartheta=d\vartheta+\Gamma\!\wedge\!\vartheta
+\vartheta\!\wedge\!\Gamma=T^\mu\gamma_\mu, \\
\Gamma:=(1/4)\Gamma^{\mu\nu}\gamma_{\mu\nu}, &\quad
R:=d\Gamma+\Gamma\wedge\Gamma=(1/4)R^{\mu\nu}\gamma_{\mu\nu}.
\end{align}
The differentials of Dirac spinors are
\begin{align}
D\psi:=d\psi+\Gamma\psi, &\qquad
D\bar\psi:=d\bar\psi-\bar\psi\Gamma, \\
D^2\psi\equiv R\wedge\psi, &\qquad
D^2\bar\psi\equiv-\bar\psi\wedge R.
\end{align}

\bigskip \noindent
\section{Some
spinor curvature identities}
\smallskip

A key to our work is certain {\em spinor-curvature identities}
(SCIs) \index{spinor-curvature identities} \cite{NTZ94}. They
readily follow from
\begin{align}
&d[\bar\psi A\wedge D(B\psi)-(-1)^a D(\bar\psi A)\wedge
B\psi]\equiv
\nonumber  \\
&2D(\bar\psi A)\wedge  D(B\psi)+(-1)^a\bar\psi A\wedge D^2(B\psi)
-(-1)^a D^2(\bar\psi A)\wedge  B\psi.
\end{align}
(Here $A$ and $B$ are Clifford algebra valued forms of rank $a$
and $b$.) Using $D^2(B\psi)=R\wedge B\psi ,$ $D^2(\bar\psi
A)=-\bar\psi A\wedge R$ we find the SCIs
\begin{align}
 2D(\bar\psi A)\wedge D(B\psi) &\equiv 2(-1)^a\bar\psi A\wedge
R\wedge (B\psi) \nonumber\\
&+ d[\bar\psi A\wedge D(B\psi)-(-1)^a D(\bar\psi A)\wedge B\psi].
\label{sci}
\end{align}
Qualitatively
$(D\psi)^2 \equiv \psi^2 R$ plus a total differential.
One
can get various linear combinations of the curvature depending on the
choice of $A,B$.
We have found three special cases with gravitational applications;
they contain, respectively,
(i) the Einstein 3-form in 4 dimensions,
(ii) the scalar curvature in 3 dimensions, and
(iii) the scalar curvature in 4 dimensions.

We know that one can also have identities of the general form
(\ref{sci}) with the spinor field $\psi$ replaced by a vector or
tensor.  However we have not found any such {\it tensor-curvature
identities} with the property that the term linear in the
curvature reduces to the Einstein or scalar curvature---which is
what we need for our gravitational applications.  This technical
point is apparently the reason why we need to use spin 1/2 to help
clarify certain things about gravity, a field which is
fundamentally spin 2.

\bigskip
\noindent
\section{The energy-momentum of gravitating systems}
\index{gravitational energy-momentum}
\medskip

Isolated gravitating systems have gravitational fields which are
asymptotically flat (very far away the field is essentially
Newtonian).  For such spaces the total energy-momentum (EM) is
well defined \cite{MTW}.  An essential fundamental theoretical
requirement (from thermodynamics and stability:  otherwise systems
could emit an unlimited amount of energy while decaying deeper
into ever more negative energy states) is that the energy of
gravitating systems should be {\it positive}.  Essentially this
means that gravity acts like a purely attractive force.  (The
total energy is just $E=Mc^2,$ with $M$ being the apparent
asymptotic Newtonian mass; thus positive energy means $M>0,$ hence
an attractive force.) This was finally rigorously  proved for GR
by Schoen and Yau \cite{SY79} via an indirect argument. Soon
thereafter Witten \cite{Wit81} gave his celebrated direct
spinorial positive energy proof (see also \cite{Nes81}).

Although positivity has been settled, the {\em location} of the
energy of gravitating systems has remained an outstanding issue
since Einstein's day.  Sources (which have a well defined local EM
density) exchange EM with the gravitational field ---{\it
locally}--- hence it was natural to expect a {\em local}
gravitational EM density.  But no suitable expression has been
found.  Standard techniques (e.g., translation symmetry and
Noether's theorem) give only non-covariant (coordinate dependent)
{\it pseudotensor} expressions (for recent discussions see
\cite{CNC99,CNC00}).  It was eventually realized that this
localization problem is entirely consistent with the {\em
equivalence principle}:  `gravity is not observable at a point'
\cite{MTW}. Nowadays the more popular idea is {\em quasilocal} EM
\index{quasilocal energy-momentum} (i.e., associated with a closed
2 surface) \cite{BY93}.

\bigskip
\noindent
\section{The Hamiltonian approach}
\index{Hamiltonian}
\medskip
A good definition of energy is: the value of the Hamiltonian.
For both a finite or infinite region
 the Hamiltonian
 includes a 3-volume term
and a bounding 2-surface integral term:
\begin{equation}
H(N)=\int_{V}N^\mu{\cal H}_\mu+\oint_{S=\partial V}{\cal B}(N),
\end{equation}
here $N$ is the spacetime vector field describing the evolution (timelike
displacement) of the spatial volume V.

For gravitating systems it follows from Noether's theorem and
local translation (diffeomorphism) symmetry that ${\cal
H}_\mu\propto \hbox{field eqns}$ (the initial value constraints).
Consequently the volume term (although it serves to generate the
Hamiltonian equations) has vanishing numerical value.  The
boundary term plays a doubly important role:  it gives the value
of the quasilocal quantities, and it also gives the boundary
conditions.  In addition to its dependence on the dynamic fields
and the displacement vector field, the boundary term generally
also depends on a choice of reference fields (which determines the
``zero'' for the quasilocal values).  There is yet considerable
freedom.  Indeed, at least formally, one could say that are an
infinite number of possible choices for the boundary term ${\cal
B};$ each corresponds to a distinct selection among the infinite
number of conceivable choices for the boundary conditions. Thus
additional criteria are very much needed.  We proposed
``covariant-symplectic'' boundary conditions;
 \index{boundary conditions!covariant-symplectic}
 it turns out
that there are only two choices that satisfy this property
(essentially they correspond to Dirichlet and Neumann boundary
conditions) \cite{CNC99,CNT95,CN99,CN00}.

For GR in terms of differential forms the standard ``ADM'' Hamiltonian
is given by the spatial integral of the 3-form
 \begin{equation}
{\cal H} = -N^\mu R^{\alpha\beta} \wedge\eta_{\alpha\beta\mu}
- i_N \Gamma^{\alpha\beta}\, D \eta_{\alpha\beta}
+ d {\cal B}(N).
\label{grh}
\end{equation}
This is easily verified if one just notes that
 $R^{\alpha\beta}\wedge\eta_{\alpha\beta\mu}\equiv-2G^\nu{}_\mu \eta_\nu$
is a 3-form version of the Einstein tensor.  When (\ref{grh}) is
integrated over space only the coefficient of $\eta_0$
contributes; that coefficient is $2G^0{}_\mu ,$ the well known
covariant components that make up the ADM Hamiltonian, see, e.g.,
\cite{MTW,IN80}.

The total differential, when integrated over a spatial region,
yields an integral over the boundary of the region.  The weak
field limit (which applies asymptotically) fixes the form of
${\cal B},$ but only to linear order.  Our best choice for the
boundary term in general is
\begin{equation}
{\cal B}(N) =
  \Delta \Gamma^{\alpha\beta} \wedge i_N \eta_{\alpha\beta} +
{\stackrel{\scriptstyle\circ}{D}}{}^{[\beta} {\stackrel
{\scriptstyle\circ}{N}}
 {}^{\alpha]} \, \Delta \eta_{\alpha\beta}  \, ,
\label{grb}
\end{equation}
where $\Delta\Gamma:=\Gamma-{\stackrel{\scriptstyle
\circ}{\Gamma}} ,$
$\Delta\eta:=\eta-{\stackrel{\scriptstyle\circ}{\eta}} .$ Here
${\stackrel{\scriptstyle\circ}{\Gamma}} ,$
${\stackrel{\scriptstyle\circ}{\eta}}$  indicate reference
values---usually taken to be the flat space field values. (Note:
all of the quasilocal quantities vanish when the dynamic fields
take on the reference values on the boundary.) The boundary term
(\ref{grb}) yields quasilocal values with good limits
asymptotically \cite{HN93,HN96} and has good correspondence with
other well established expressions \cite{CN99}.

The significance of the choice of boundary term (\ref{grb}) is
revealed by the variation of the Hamiltonian:
\begin{equation}
\delta{\cal H}(N)=\delta\vartheta^\alpha\wedge {\frac{\delta{\cal
H}(N)}{\delta\vartheta^\alpha}}+ \delta\Gamma^{\alpha\beta}\wedge
{\frac{\delta{\cal H}(N)}{\delta\Gamma^{\alpha\beta}}}
+di_N(\Delta\Gamma^{\alpha\beta}\wedge
\delta\vartheta^\mu\wedge\eta_{\alpha\beta\mu}).
\end{equation}
In addition to the field equation terms (we do not need their
explicit functional form here) we obtain a Hamiltonian boundary
variation term. It has a {\em symplectic
structure}\index{symplectic structure} \cite{KT79} which,
according to the {\em boundary variation principle}\index{boundary
variation principle}, reveals which variables are to be held
fixed. In this case we should fix (certain projected components
of) the orthonormal frame $\vartheta^\mu ,$ (geometrically that is
equivalent to holding the metric fixed) \cite{CNT95,CN99,CN00}.

\bigskip
\noindent
\section{ Spinor expressions}

\medskip
Using certain special cases of the general spinor curvature
identity (\ref{sci}), we obtain three alternate spinor
formulations for the GR Hamiltonian and its boundary term.  From
each we get both a positive energy proof and a quasilocal EM
expression.  We discuss the first case in some detail and then
briefly survey the novel features in the other two cases.

\medskip
\noindent
\subsection
{The Witten spinor approach}
 \index{Witten's positive energy proof}

\medskip
In the above ADM Hamiltonian 3-form (\ref{grh}), use a spinor
parameterization for the Hamiltonian displacement
 \begin{equation}
N^\mu ={\overline\psi}\gamma^\mu\psi.
 \end{equation}
With an appropriately adjusted boundary
term, a suitable {\it spinor-curvature identity}
then gives the Hamiltonian 3-form
associated with the famous Witten positive energy proof:
\begin{align}
{\cal H}_{w}(\psi) &:=
 4 D{\overline\psi}\wedge \gamma \vartheta \wedge D \psi
- i_N\Gamma^{\alpha\gamma}\,  D \eta_{\alpha\gamma} \nonumber\\
&\equiv
 - N^\mu R^{\alpha\gamma}\wedge\eta_{\alpha\gamma\mu}
-i_N\Gamma^{\alpha\gamma}\,  D \eta_{\alpha\gamma}
 + d{\cal B}_{w}\, ,
\label{wh}
\end{align}
where
\begin{equation}
{\cal B}_{w}:=2( {\overline \psi}
\gamma \vartheta \wedge D \psi
   + D {\overline \psi} \wedge\gamma \vartheta \psi)\, .
\label{wb}
\end{equation}
We stress that this is an acceptable alternate form for (\ref{grh}),
the GR Hamiltonian
\cite{Nes84}.
Note that the spinor field can take on almost any value, as long as it
is asymptotically constant.  For such spinor fields the boundary term
(\ref{wb}), notwithstanding appearances, actually gives the same
asymptotic values as those given by (\ref{grb}).

For positive energy, first note that, because of vanishing torsion,
one of the Hamiltonian terms vanishes:
 $D\eta_{\mu\nu}=T^\lambda\wedge\eta_{\mu\nu\lambda}=0$.
The Hamiltonian density is thus
\begin{equation}
{\cal H}_{w}(\psi) =
4 \left(D_\alpha{\overline\psi} \gamma \gamma_\lambda  D_\beta \psi\right)
 \vartheta^\alpha\wedge
 \vartheta^\lambda\wedge
 \vartheta^\beta
 \equiv
4 \left(D_\alpha{\overline\psi} \gamma \gamma_\lambda  D_\beta \psi\right)
 \eta^{\alpha\lambda\beta\mu}\eta_\mu.
\end{equation}
Now, using the grade 3 identity
\begin{equation}
\gamma_\lambda
\gamma_{\mu\nu}+\gamma_{\mu\nu}\gamma_\lambda\equiv2\gamma_{\lambda\mu\nu}
\equiv 2\eta_{\lambda\mu\nu\kappa}\gamma\gamma^\kappa,
\end{equation}
the Hamiltonian density becomes
\begin{equation}
{\cal H}_{w}(\psi)= 2D_\alpha{\overline\psi}\left(
\gamma^{\alpha\beta}\gamma^\mu+\gamma^\mu\gamma^{\alpha\beta}\right)
D_\beta\psi \eta_\mu.
\end{equation}
When integrated over space only the $\mu=0$ term survives; this
means that the other two indices must be spatial.  Finally, using
$\gamma^{ab}\equiv\gamma^a\gamma^b-g^{ab}$,  $g^{ab}=-\delta^{ab}$
and the usual type of representation wherein $\beta=\gamma^0$, the
Hamiltonian density has the 3+1 (space+time) decomposition
 \begin{equation}
{\cal H}_w(\psi)\simeq
2D_a{\overline\psi}\left(
\gamma^{ab}\gamma^0+\gamma^0\gamma^{ab}\right)
D_b\psi \eta_0 \,
\propto  \,  |D_k\psi|^2-|\gamma^k D_k\psi|^2.
\end{equation}
Now, exploiting the freedom in the choice of $\psi$, note that the
Hamiltonian density (and consequently the energy) is manifestly
non-negative, for any $\psi$ solving the (3-dimensional, elliptic)
{\it Witten equation}\index{Witten spinor equation}: $\gamma^k
D_k\psi=0$.  (This proof, due to Witten, was later derived
directly from the classical, {\em anti-commuting} spinor,
supergravity result \cite{HS83,Des83}; there is an interesting
argument that GR has positive energy {\em because} it admits a
supersymmetric extension.)

Moreover we get a `locally positive localization' for
gravitational energy (albeit the {\em localization} actually
depends on the solution of an elliptic equation and hence really
depends on the fields globally). Examined in more detail, the
argument also shows, (i) that the 4 energy-momentum $P^\mu$ is
future time-like, and (ii) $P^\mu$ vanishes only for Minkowski
space.

Altogether these are very beautiful arguments for some nice
important results. Nevertheless, more than 20 years later, it is
still not clear as to how much the Witten argument really captures
the correct physics.

Consider the (static, spherically symmetric) Schwarzschild
solution in an isotropic Cartesian frame ($\vartheta^0=Ndt ,$
$\vartheta^i=\varphi^2 dx^i ,$ with $\varphi=1+m/2r ,$
$N\varphi=1-2m/r$).  The Witten equation is easily solved:
$\psi=\varphi^{-2}\psi_{const} .$  One can then substitute this
solution into the Hamiltonian and the boundary term, and thereby
conclude that, for the Schwarzschild solution, 1/8 of the energy
has been ``localized'' within the black hole horizon and 7/8 is
outside.  We have no physical understanding of this curious
distribution.

Note also that for closed spaces, such as an $S^3$ type
cosmology, the spatial hypersurface has no
boundary, so it should have vanishing total energy.  Hence the Witten
positive energy proof (or indeed any other positivity proof) should
not go through.  We are not yet sure which step in the Witten argument
breaks down for such closed spaces.

The Witten spinor Hamiltonian is also important for the quasilocal
values it can yield.  When integrated over a finite spatial region,
the Hamiltonian boundary term ${\cal B}_w$ (\ref{wb}) defines a
quasilocal energy-momentum for any choice of the spinor field on the
boundary.  This is a popular approach to quasilocal energy.  Several
similar quasilocal boundary expressions of this type have been
investigated.  In this case one wants to determine the spinor field
quasilocally (i.e., it should depend on the fields only on the
boundary 2-surface $S$).  In particular Dougan and Mason \cite{DM91}
take $\psi$ to be ``holomorphic'' (satisfying the 2-dimensional elliptic
equation $\gamma^A\nabla_A\psi=0$)
on the boundary.  A nice investigation of the various options for
equations to select the value of the spinor field for such quasilocal
expressions has been carried out by Szabados \cite{Sza94}.

\bigskip
\noindent
\subsection{An SU(2) formulation}
\medskip

Here we briefly describe a 3-dimensional alternative, similar to
but distinct from the Witten formulation, which uses the SU(2)
spinors of the 3-dimensional spatial hypersurface \cite{NT94}.
  We begin from the well known ADM Hamiltonian \cite{MTW,IN80}
\begin{align}
H(N) =& \int d^3 x \Bigl\{ N\bigl[
g^{-{\frac{1}{2}}}(\pi^{mn}\pi_{mn}-{\frac{1}{2}}\pi^2)
-g^{\frac{1}{2}}R\bigr]+2\pi^m{}_k\nabla_m N^k\Bigr\} \nonumber \\
& + \oint dS_k N \delta^{kc}_{am}g^{mb}\Gamma^a{}_{bc},
\label{admham}
\end{align}
which we have supplemented by boundary term expressions that are valid
in asymptotic Cartesian frames.  In this case we concern ourself only
with energy (not momentum), so we take the shift
$N^k$ to vanish.  Within the ADM Hamiltonian density the scalar
curvature term and the boundary term do not have a definite sign; the
idea is to replace them with alternatives that are more definite.  The
3-scalar curvature can be replaced using
$N=\varphi^{\dag}\varphi$ and the 3-dimensional SCI
\begin{equation}
2\Bigl[\nabla(\varphi^{\dag}i\sigma)\wedge\nabla\varphi-
\nabla\varphi^{\dag}\wedge\nabla(i\sigma\varphi)\Bigr]
\equiv dB_{su2} -
\varphi^{\dag}\varphi R^{ab}\wedge\epsilon_{abc}\vartheta^c,
\end{equation}
where $R^{ab}\wedge\epsilon_{abc}\vartheta^c=Rg^{1/2} d^3x ,$
$\sigma:=\sigma_a\vartheta^a$ is a Pauli matrix valued one form,
and
\begin{equation}
B_{su2}:=2 [\nabla\varphi^{\dag}\wedge
i\sigma\varphi+\varphi^{\dag}i\sigma\wedge\nabla\varphi],
\label{SU2hb}
\end{equation}
is a legitimate (since they agree asymptotically) alternative to the
boundary term in (\ref{admham}).

The Hamiltonian density now takes the form
\begin{align}
{\cal H}(\varphi)=& (\varphi^{\dag}\varphi)g^{-{\frac{1}{2}}}
\Bigl[ \pi^{mn}\pi_{mn} - {\frac{1}{2}}\pi^2\Bigr]d^3x \nonumber\\
&+\Bigl[\nabla(\varphi^{\dag}i\sigma)\wedge\nabla\varphi-
\nabla\varphi^{\dag}\wedge\nabla(i\sigma\varphi)\Bigr].
\end{align}
By arguments similar to those used in connection with the Witten
Hamiltonian, the quadratic $\nabla\varphi$ terms can be
diagonalized to the form
$|\nabla_k\varphi|^2-|\sigma^k\nabla_k\varphi|^2 .$ Consequently
the Hamiltonian density is non-negative---and hence the total
energy is positive--- on {\it maximal} spacelike hypersurfaces (in
this standard time gauge condition the trace of the ADM canonical
momentum $\pi$  vanishes) if the 3-dimensional spinor is chosen to
satisfy the 3-dimensional elliptic equation
$\sigma^k\nabla_k\varphi=0 .$

Again, the boundary term ${\cal B}_{su2}$ gives a quasilocal
energy for any choice of $\varphi ;$ in particular, as in the
previous case, we can use  holomorphic spinors satisfying
$\sigma^A\nabla_A\varphi=0 .$

\bigskip
\noindent
\subsection{ The QSL approach}
\medskip

Our last alternative has some similar features to the above two
approaches but differs from them in a very fundamental way.  In
the above the spinor field was introduced into the Hamiltonian as
a technical device to aid in obtaining a locally non-negative
Hamiltonian density.  Instead one can introduce a spinor field
into the Lagrangian, then it becomes a basic dynamical
gravitational field. The key is another {\it spinor-curvature\/}
identity, this time involving the 4-dimensional scalar curvature,
which led us to a {\it Quadratic Spinor Lagrangian} (QSL)
\index{Quadratic Spinor Lagrangian} for GR \cite{NT95}.  (For the
relation to teleparallel GR, aka TEGR, GR${}_{||} ,$ see
\cite{TN99}) The Einstein-Hilbert scalar curvature Lagrangian
equals (up to an exact differential) the QSL
\begin{equation}
{\cal L}_{qs}:=2D{\overline \Psi}\gamma \wedge D\Psi
\equiv -R*1 + d(D{\overline \Psi}\wedge\gamma \Psi
+{\overline \Psi}\gamma \wedge D\Psi),
\end{equation}
where $\Psi=\vartheta\psi$ is a spinor one-form field. The spinor
field is to be varied subject to the normalization constraints
$\bar\psi\psi=1 ,$ $\bar\psi\gamma\psi=0$ (which can be enforced
via Lagrange multipliers).

In order to construct the Hamiltonian it is convenient to work from
the corresponding first order Lagrangian:
\begin{equation}
{\cal L}_{\Psi}:=D{\overline\Psi}\wedge P+{\overline P}\wedge
D\Psi+ {\textstyle{\frac{1}{2}}}{\overline P}\wedge\gamma P\,
\end{equation}
(which yields the pair of first order equations:  $2D\Psi=-\gamma P$
and its conjugate).  From this we find the covariant Hamiltonian
3-form
\begin{align}
{\cal H}_{\Psi} :=& {\overline P}\wedge\Lie_N\Psi+\Lie_N{\overline
\Psi}\wedge P
-i_N{\cal L}_{\Psi} \nonumber\\
\equiv& -i_N({\textstyle{\frac{1}{2}}}{\overline P}\wedge\gamma P) \nonumber\\
&-\left[i_N{\overline \Psi}DP+D{\overline \Psi}\wedge i_N P
+{\overline\Psi}\wedge i_N\omega P\right. 
-
\left. d(i_N{\overline \Psi}P)+ \hbox{c.c.}\right] \, ,
\end{align}
(the time derivative here is given by the Lie derivative).
 The Hamiltonian boundary term
\begin{equation}
{\cal B}(N)=i_N{\overline\Psi}P + {\overline P}i_N\Psi\, ,
\label{QSLhb}
\end{equation}
again yields quasilocal values for any choice of spinor field on
the boundary.

One distinguishing feature of this approach, in which the spinor
field is introduced into the Lagrangian, is that the displacement
vector field $N$ remains entirely independent of the spinor field.
Another is that the connection does not appear as a primary
dynamic variable. Nevertheless it is again possible to arrange
that the Hamiltonian is locally non-negative.  By arguments
similar to those used before, with the lapse-shift choice
$N^\mu=(f^2,0,0,0) ,$ and the spinor field satisfying the
(3-dimensional, elliptic) conformal Witten equation: $\gamma^k
D_k(f\psi)=0 ,$ the QSL Hamiltonian density is locally
non-negative on maximal slices.  This yields another locally
positive `localization' along with a positive energy proof.

For further discussion of additional details and features of these
cases see the already cited references as well as the overview in
\cite{CNC00}.

\bigskip
\noindent
\section{ Properties of the spinor expressions}

The spinor quasilocal expressions have certain common properties.
A noteworthy one is that the expressions are essentially algebraic
in $N^\mu ;$ unlike the standard GR expression (\ref{grb}) they
have no $DN$ terms.  (The importance of this feature will be
discussed soon.) Another is perhaps the real {\it beauty} of these
spinor formulations:  they do not need an {\it explicit} reference
configuration---whereas in the standard GR expression (\ref{grb})
the reference configuration is essential.  When examined in detail
it can be seen that the spinor field {\it implicitly} determines
the reference values \cite{CNT95,CNC00}.  That is one of its jobs.

There are other distinct roles for the spinor fields.  The {\it
Hamiltonian variation boundary principle} tells us what must be
held fixed on the boundary.  This clarifies the role of the
fields. Essentially for the spinor formulations we find that (in
addition to fixing the usual frame components) the spinor field
$\psi$ should be held fixed on the boundary.  In each of the cases
this has a distinct significance.  For the Witten Hamiltonian
fixing the spinor field fixes the displacement $N^\mu ,$ for the
SU(2) spinor alternative it means holding the {\it lapse} fixed,
whereas for the QSL fixed $\psi$ amounts to fixing the observer's
orthonormal frame on the boundary.  In each case the spinor field
has a different role.  Altogether we have good examples of what
spinors can mean geometrically and physically.

In each case, for displacements corresponding to 4-dimensional
space-time translations, the quasilocal spinor expressions
(\ref{wb},\ref{SU2hb},\ref{QSLhb}) asymptotically agree with a
standard one (\ref{grb}) and hence each expression gives similar
reasonable values for the total energy-momentum.

\bigskip
\noindent
\section{ Angular Momentum and Center-of-Mass Moment}
 \index{angular momentum}
 \index{center-of-mass}
\smallskip

However energy-momentum is not the whole story.  For an asymptotically
flat space each asymptotic symmetry has its associated conserved
quantities.  Energy-momentum is associated with spacetime
translations.  Such spaces also have asymptotic rotation and Lorentz
boost symmetry.  A proper physical Hamiltonian formalism allows for
displacements which have the asymptotic Poincar{\'e} form
\begin{equation} N^\mu=N^\mu_\infty+\lambda^\mu{}_\nu x^\nu.
\label{poinasyt}
 \end{equation}
 Here $N^\mu_\infty$ is a constant
spacetime translation and the constants
$\lambda^{\mu\nu}=\lambda^{[\mu\nu]}$ describe an infinitesimal Lorentz
transformation (including rotations and boosts).  The value of the
Hamiltonian generating the Lorentz displacements then gives 6
additional conserved quantities \cite{RT74,BO87}.
The physical significance of these additional quantities is most
easily recognized by recalling that for
a relativistic particle
\begin{equation}
L^{\mu\nu}:=x^\mu p^\nu-x^\nu p^\mu
\end{equation}
includes both {\em angular momentum},
\begin{equation}
L^{ij}:=x^i p^j-x^j p^i, \quad \hbox{and}\quad
L^{0k}:=x^0 p^k-x^k p^0,
\end{equation}
the {\it center-of-mass moment}.

The conventional variable GR Hamiltonian description of  section 5
does yield good values for these additional quantities for
gravitating systems.  An important contribution here is the
$i_N\Gamma\sim DN$ connection-M{\o}ller-Komar type term in
(\ref{grb}) (such terms have often been overlooked in quasilocal
investigations \cite{BY93}).  We found that this term plays a key
role in: (i) black hole thermodynamics \cite{CN99}, (ii) certain
angular momentum calculations \cite{HN93,HN96,Vu00} (iii) {\em
all} center-of-mass moment calculations \cite{Mng02}. (Note: from
(\ref{poinasyt}) it follows that asymptotically the $DN$ term has
a contribution proportional to $\lambda$.)

Now such terms are completely absent in all three of our spinor
formulations.  Without them we, (i) do not know how to obtain the
first law of black hole thermodynamics, (ii) have difficulties in
obtaining the angular momentum (within the Witten Hamiltonian
(\ref{wh}), modifying the displacement to
$\bar\psi\gamma\gamma^\mu\psi$ will work---but only {\em if} we
take the strange asymptotics $\psi^{\dag}\psi\sim r$), (iii)
simply {\em cannot see how} to get the correct center-of-mass
moment. Now we have not entirely given up trying, but so far we
have not managed to include the effects of such $D^{[\mu}N^{\nu]}$
type terms in the spinor formulations.  Presently we are preparing
detailed discussions of these crucial issues.  Here we can
summarize our tentative conclusions but regretfully cannot include
any more of the technical supporting details.

\bigskip
\noindent
\section{Conclusions}
\medskip
We considered the spinor formulations for the Hamiltonian and the
associated positive energy proofs and quasilocal expressions.  At
first the role of the spinor field seemed mysterious.  However the
boundary variation principle clarifies the role of the spinor
field (and indeed all other variables).  Our spinor Hamiltonian
expressions illustrate the variety of roles that spinor fields can
play. Spinors give beautiful positive energy proofs (especially
Witten's) and very neat formulas for quasilocal energy-momentum.
It is especially noteworthy that there is no need for extra
reference fields on the boundary.

Yet it seems that such spinor formulations have a serious
limitation: the present expressions cannot give angular momentum
and the center-of-mass moment.  Moreover, we do not see how any
natural adjustment of these spinor formulations for quasilocal
Hamiltonian boundary terms can successfully give expressions for
these quantities.  Apparently the spinor Hamiltonians cannot give
all of the physically conserved quantities of asymptotically flat
spacetimes. Hence, although they are deservedly popular and quite
good for many purposes, we now believe that such spinor
formulations cannot really capture {\it in a physically correct
way} the still elusive gravitational energy-momentum.

\bigskip
\section*{Acknowledgments}
The work was supported by the National Science Council of the Republic
of China under the grants NSC90-2112-M-008-041, NSC90-2112-M-002-055.


\small \vskip 1pc {\obeylines
 \noindent Chiang-Mei Chen
 \noindent Department of Physics
 \noindent National Taiwan University
 \noindent Taipei, 106, Taiwan, R.O.C.
 \noindent E-mail: cmchen@phys.ntu.edu.tw
 \vskip 1pc
  \noindent James M. Nester
 \noindent Department of Physics
 \noindent National Central University
 \noindent Chung-Li, 320, Taiwan, R.O.C.
 \noindent E-mail: nester@phy.ncu.edu.tw
 \vskip 1pc
  \noindent Roh-Suan Tung
 \noindent Department of Physics,
 \noindent National Central University
 \noindent Chung-Li, 320, Taiwan, R.O.C.
 \noindent E-mail: rohtung@phy.ncu.edu.tw
} \vskip 6pt

\printindex

\bigskip


\begin{thebibliography}{99}
\def\topsep{0pt}
\def\parsep{0pt plus 5pt minus 1pt}
\def\itemsep{-0.5ex} 
\small               
%

\bibitem{ADM62} R. Arnowitt, S. Deser  and C. W. Misner,
 in {\it Gravitation: an Introduction to Current Research},
Ed. L. Witten, Wiley, New York, 1962, 227--265.

\bibitem{BO87} R. Beig and N. {\sc \'O} Murchadha,
{\it Ann. Phys. \bf 174} (1987), 463--498.

\bibitem{BY93} J. D. Brown and J. W. York, Jr., {\it Phys.\ Rev.\ D
}{\bf 47} (1993), 1407--1409.

\bibitem{CNC99}
    C.-C. Chang, J. M. Nester and C.-M. Chen,
    {\it Phys.\ Rev.\ Lett.\ \bf 83} (1999), 1897--1901;
    {arXiv: gr-qc/9809040}.

\bibitem{CNC00}
    C.-C. Chang, J. M. Nester and C.-M. Chen,
    in {\it
    Gravitation and Astrophysics}, Eds. Liao Liu, Jun Luo, X-Z. Li, J-P. Hsu, World Scientific, Singapore, 2000,
163--173; {arXiv: gr-qc/9912058}.

\bibitem{CN99}
    C.-M. Chen and J. M. Nester,
    {\it Class.\ Quantum\ Grav.\ \bf 16} (1999), 1279--1304;
    {arXiv: gr-qc/9809020}.

\bibitem{CN00}
    C.-M. Chen and J. M. Nester,
 {\it Gravitation \& Cosmology \bf 6} (2000), 257--270;
      {arXiv: gr-qc/0001088}.

\bibitem{CNT95}
C. M. Chen,  J. M. Nester and R. S. Tung, {\it Phys.\ Lett.\ \bf A
203} (1995), 5--11.

\bibitem{Des83}
S. Deser, {\it Phys.\ Rev.\ \bf D 27} (1983), 2805--2808.

\bibitem{DT77}
S. Deser and C. Teitelboim, {\it Phys.\ Rev.\ Lett.\ \bf 39}
(1977), 249--252.

\bibitem{DMH91} A. Dimakis and F. M{\"u}ller-Hoissen,
{\it Class.\ Quantum Grav.\ \bf 8} (1991), 2093--2132.

\bibitem{DM91} A. Dougan and L. Mason, {\it Phys. Rev. Lett. \bf 67} (1991),
2119--2122.

\bibitem{Est91} F. Estabrook, {\it Class. Quantum Grav. \bf 8} (1991), L151--154.

\bibitem{Gri78}
M. T. Grisaru, {\it Phys.\ Lett.\ \bf 73B} (1978), 207--208.

\bibitem{HN93} R. D. Hecht and J. M. Nester, {\it Phys. Lett. A \bf 180} (1993),
324--331.

\bibitem{HN96} R. D. Hecht and J. M. Nester, {\it Phys. Lett. A \bf 217} (1996),
81--89.

\bibitem{HS83}
G. T. Horowitz and A. Strominger, {\it Phys.\ Rev.\ \bf D 27}
(1983), 2793--2804.

\bibitem{IN80} J. Isenberg and J. M. Nester, Canonical Gravity, in:
{\it General Relativity and Gravitation: One Hundred Years After
the Birth of Albert Einstein} Vol I., Ed. A. Held, Plenum, New
York, 1980, 23--97.

\bibitem{KT79} J. Kijowski and W. M. Tulczyjew, {\it A
Symplectic Framework for Field Theories (Lecture Notes in Physics}
vol 107), Springer, Berlin, 1979.

\bibitem{Mng02} F. F. Meng, Quasilocal center-of-mass moment in GR,
MSc.
Thesis, National Central University, 2002 (unpublished).

\bibitem{Mi87} E. W. Mielke,
{\it Geometrodynamics of Gauge Fields---on the Geometry of
Yang-Mills
 and Gravitational Gauge Theories}, Akademie, Berlin, 1987.

\bibitem{MTW} C. W. Misner, K. Thorne  and J. A. Wheeler, {\it
Gravitation} Freeman, San Fransisco, 1973.

\bibitem{Nes81} J. M. Nester, {\it Phys. Lett. A \bf 83} (1981), 241--242.

\bibitem{Nes84} J. M. Nester, The gravitational Hamiltonian,
in {\it Asymptotic Behavior of Mass and Space-Time Geometry
(Lecture Notes in Physics} vol 202), Ed. F. Flaherty, Springer,
Berlin, 1984, 155--163

\bibitem{Nes88} J. M. Nester, {\it Class.\ Quantum Grav. \bf 5} (1988),
1003--1010.

\bibitem{NT94} J. M. Nester and R. S. Tung, {\it Phys. Rev. D \bf
49} (1994), 3958--3962.

\bibitem{NTZ94} J. M. Nester, R. S. Tung and V. Zhytnikov, {\it Class. Quant.
Grav. \bf 11} (1994), 983--987.

\bibitem{NT95}
J. M. Nester and R. S. Tung, {\it Gen.\ Rel.\ Grav.} {\bf 27}
(1995), 115--119.

\bibitem{RT74} T. Regge and C. Teitelboim, {\it Ann. Phys.
\bf88} (1974), 286--319.

\bibitem{SY79} R. Schoen and S. T. Yau, {\it Comm. Math Phys. \bf 65} (1979), 45--76.

\bibitem{Sza94} L. Szabados, {\it Class. Quantum Grav. \bf 11} (1994),
1833--1847.

\bibitem{TN99} R. S. Tung and J. M. Nester,
{\it Phys.\ Rev.\ D \bf 60} (1999),  021501.

\bibitem{Vu00} K. H. Vu, Quasilocal energy-momentum and angular momentum for
teleparallel gravity, MSc. Thesis, National Central University,
2000 (unpublished).

\bibitem{Wit81} E. Witten, {\it Comm. Math. Phys. \bf 80} (1981), 381--402.

\end{thebibliography}
\end{document}